\documentstyle[12pt]{article}
\input epsf
\begin{document}

\begin{titlepage}
\begin{center}
{\bf POSITIVITY OF NLO SPIN-DEPENDENT PARTON DISTRIBUTIONS} \\
{ C. Bourrely$^{1,2}$, E. Leader$^2$, \underline{O.V. Teryaev}$^3$}\\
{\it (1) Centre de Physique Th\'eorique - CNRS - Luminy,
Case 907 \\ F-13288 Marseille Cedex 9 - France \\
(2)Birkbeck College, Malet Street, London WC1E 7HX, UK\\
(3) Joint Institute for Nuclear Research (JINR), 141980 Dubna, Russia}
\end{center}
\begin{abstract}
We discuss the positivity  of the hadron density matrix in QCD.
This basic property is shown to be preserved by QCD evolution, provided the
relation $|\Delta P_{ij}(z)| \leq P_{ij}$ is valid for all kernels
for $z < 1$, and the usual "$+$" prescription is used. We comment on
the positivity restrictions for the choice of the NLO factorization scheme
for the evolution of the spin-dependent parton distributions.

\end{abstract}


\noindent Number of figures : 1

\smallskip

\noindent December 1997

\noindent (1) Unit\'e Propre de Recherche 7061

\noindent CPT-97/P.3581

\noindent Invited talk to the VII Workshop on High Energy Spin Physics
(SPIN-97) \\Dubna July 7-12

\noindent Web address: www.cpt.univ-mrs.fr
\end{titlepage}

The positivity of spin dependent parton densities is a basic property,
which allows a self-consistent partonic interpretation.

The case of next-to-leading order (NLO) is especially interesting,
as the parton distribution is no longer a directly observable quantity,
moreover it depends on the choice of factorization scheme.
For the spin-dependent case, there is also an extra ambiguity due to
the choice of $\gamma_5$
prescription. Positivity  may be considered as an extra
constraint, restricting this ambiguity. The present paper is devoted to
a first approach to this problem.

The current next-to-leading (NLO)
parametrizations \cite{GS,GRSV} are chosen in such a way, that  positivity
for all helicity parton distributions is respected at the initial value
$Q_0^2$, i.e.
$|\Delta f (z, Q^2)| \leq f (z, Q^2)$.

One may wonder, to what extent the $Q^2$ evolution is compatible with
positivity. The answer becomes
clear when one  interprets
the QCD evolution as a kinetic flow in $x$-space.

In its standard form \footnote{For brevity, the
argument $t$ will not be written down explicitly in the parton densities.}
(for the time being, we shall confine ourselves to the
unpolarized non-singlet
(NS) case)
\begin{eqnarray}
{dq(x) \over{dt}}=
{\alpha_s \over {2 \pi}} \int_x^1 dy {{q(y)} \over y} P ({x \over y}),
\label{GLAP}\end{eqnarray}
may be interpreted as a 'time' $t=ln Q^2$  evolution  of the
'particles' density $q$ in the one dimensional space
$0 \leq x \leq 1$ due to the flow from the right to the left,
with the probability per unit time equal to the splitting kernel $P$.
The key element in such an interpretation is the problem of the
infrared (IR) singular terms in $P$, which was considered in detail
some years ago \cite{CQ} (see also \cite{DP}).
The kinetic interpretation is preserved
provided the $'+'$ form of the kernel is written in the following way
\begin{eqnarray}
P_+ (z) =P(z)-\delta(1-z)\int_0^1 P(y) dy,
\label{+}\end{eqnarray}
leading to the corresponding expression for the evolution equation
\begin{eqnarray}
{dq(x) \over{dt}}=
{\alpha_s \over {2 \pi}}[ \int_x^1 dy {{q(y)} \over y}
P ({x \over y})- q(x) \int_0^1 P(z) dz].
\label{+e}\end{eqnarray}

The negative second term in (\ref{+e}) cannot change the sign of the
distribution
because it is 'diagonal' in $x$, which means that it is proportional
to the function at the same point $x$, as on the l.h.s..
Thus when the distribution gets too close to zero,
its stops causing a decrease. This is true for both $'+'$ and $\delta(1-z)$
terms, for any value of their coefficient (if it is positive,
it will reinforce the positivity of the distribution).

Let us consider now the spin-dependent case. For simplicity, we
postpone the discussion of quark-gluon mixing for a moment, but allow
the spin-dependent and spin-independent kernels to be different,
as they are at NLO.
It is most convenient to write down the equations for definite parton
helicities, which was actually the starting point in deriving
the equations for the spin-dependent quantities \cite{AP}.
Although the form, which we shall use, mixes the contributions of
different helicities, it makes the positivity properties especially
clear. So we have
\begin{eqnarray}
{dq_+(x) \over{dt}}=
{\alpha_s \over {2 \pi}} (P_{++} ({x \over y}) \otimes q_+(y)+
P_{+-} ({x \over y}) \otimes q_-(y)), \nonumber \\
{dq_-(x) \over{dt}}=
{\alpha_s \over {2 \pi}} (P_{+-} ({x \over y}) \otimes q_+(y)+
P_{++} ({x \over y}) \otimes q_-(y)).
\label{h}\end{eqnarray}

Here $P_{++}(z)=(P(z)+\Delta P(z))/2, P_{+-}(z)=(P(z)-\Delta P(z))/2$
are the evolution kernels for definite helicities and we have used the fact
that $P_{++}=P_{--}$ and $P_{+-}=P_{-+}$, and a shorthand
notation for the convolution is adopted.
As long as $x < y$, the positivity of the initial distri\- butions
($q_+(x, Q_0^2), q_-(x, Q_0^2) \geq 0$, or
$|\Delta q (x, Q_0^2)| \leq q (x,Q_0^2)$)
is preserved, if both kernels
$P_{++},P_{+-}$ are positive, which is true, and if
\begin{eqnarray}
|\Delta P (z)| \leq P (z), \ z < 1.
\label{ineq}\end{eqnarray}

The singular terms at $z=1$  do not altering positivity,
because they appear only in the diagonal (now in helicities)
kernel $P_{++}$ (only forward scattering is
IR dangerous). From the kinetic interpretation again
the distributions $q_+,~q_-$ stop decreasing, as soon as they
are close to changing sign.

Now to extend the proof to quark gluon mixing is trivial.
One should write down the expressions for the evolutions of quark
and gluon distributions of each helicity
\begin{eqnarray}
{dq_+(x) \over{dt}}=
{\alpha_s \over {2 \pi}} (P_{++}^{qq} ({x \over y}) \otimes q_+(y)+
P_{+-}^{qq} ({x \over y}) \otimes q_-(y))  \nonumber \\
+P_{++}^{qG} ({x \over y}) \otimes G_+(y)+
P_{+-}^{qG} ({x \over y}) \otimes G_-(y),
\nonumber \\
{dq_-(x) \over{dt}}=
{\alpha_s \over {2 \pi}} (P^{qq}_{+-} ({x \over y}) \otimes q_+(y)+
P^{qq}_{++} ({x \over y}) \otimes q_-(y)) \nonumber \\
+P_{+-}^{qG} ({x \over y}) \otimes G_+(y)+
P_{++}^{qG} ({x \over y}) \otimes G_-(y),  \nonumber \\
{dG_+(x) \over{dt}}=
{\alpha_s \over {2 \pi}} (P_{++}^{Gq} ({x \over y}) \otimes q_+(y)+
P_{+-}^{Gq} ({x \over y}) \otimes q_-(y) \nonumber \\
+P_{++}^{GG} ({x \over y}) \otimes G_+(y)+
P_{+-}^{GG} ({x \over y}) \otimes G_-(y)),  \nonumber \\
{dG_-(x) \over{dt}}=
{\alpha_s \over {2 \pi}} (P_{+-}^{Gq} ({x \over y}) \otimes q_+(y)+
P_{++}^{Gq} ({x \over y}) \otimes q_-(y) \nonumber \\
+P_{+-}^{GG} ({x \over y}) \otimes G_+(y)+
P_{++}^{GG} ({x \over y}) \otimes G_-(y)).
\label{hs}\end{eqnarray}
If the inequality (\ref{ineq}) was valid for each type of partons,
\begin{eqnarray}
|\Delta P_{ij} (z)| \leq P_{ij} (z), \ z < 1; \ i,j = q,G,
\label{ineqs}\end{eqnarray}
all the kernels, appearing in the r.h.s. of such a system, are
positive. Concerning the singular terms, they are again diagonal,
now in parton type, and do not affect positivity.
The validity of these equations in LO comes just from the
way they were derived, since the (positive) helicity-dependent
kernels were in fact  calculated first in ref.\cite{AP}.

But the situation in NLO is more peculiar.
Even the spin-averaged quantities do not respect
positivity. The most striking example is the gluonic kernel
$P^{GG}$ which is negative at low $x$. This is, first of all, a signal,
that the NLO contribution comes with a positive (and large) LO one.
Moreover, for low x resummation effects are important,
coming from the most singular terms at all orders.

\begin{figure}
\begin{center}
\epsfxsize=12cm
\epsfysize=18cm
\centerline{\epsfbox{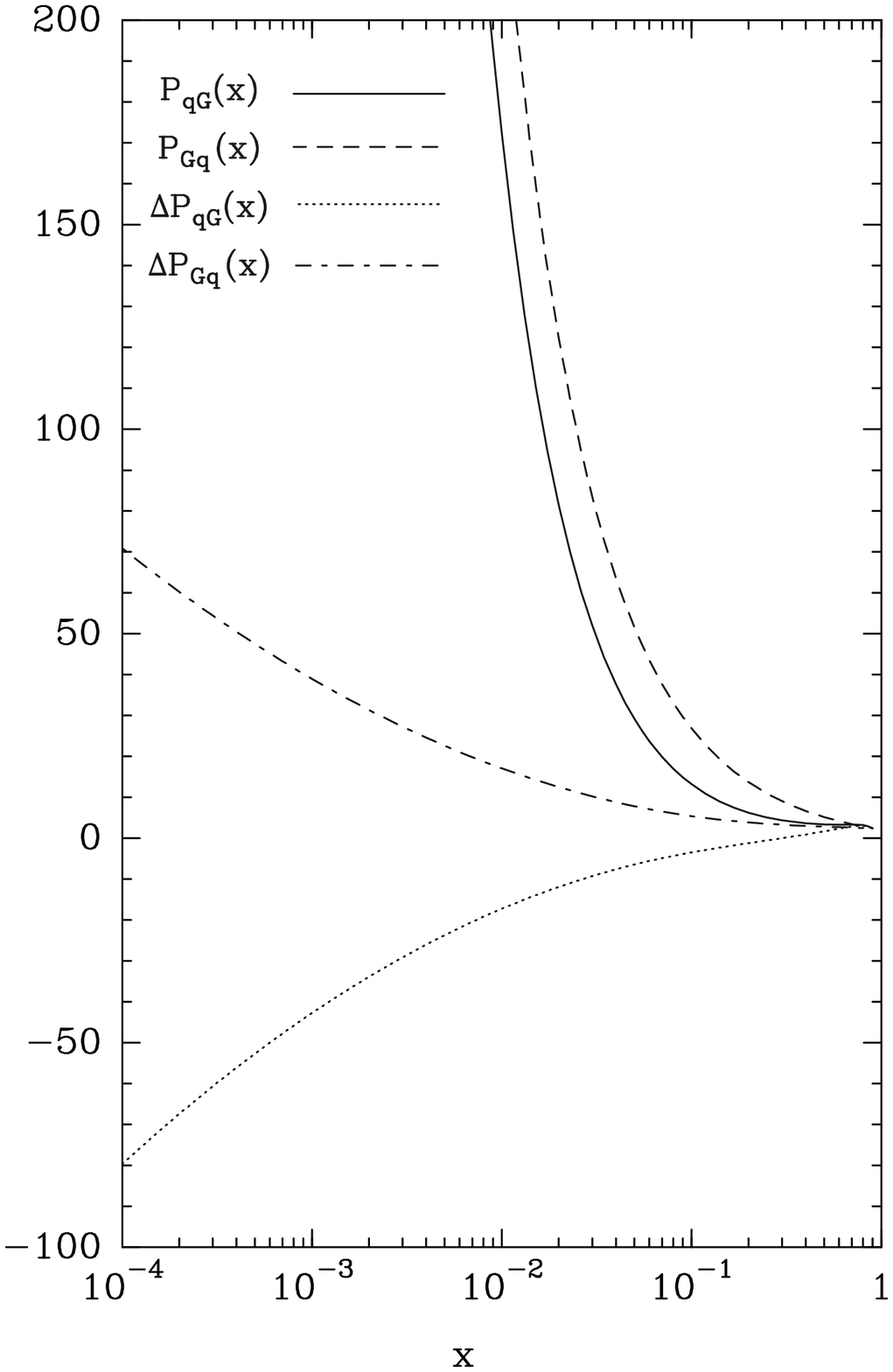}}
\caption{Unpolarized and polarized kernels as a function of $x$ including
LO and NLO contributions.}
\label{fi:dpqg}
\end{center}
\end{figure}

We performed a systematic comparison of polarized and unpolarized
NLO singlet kernels (see Figure~\ref{fi:dpqg}).

The result may be described as follows. For quark-quark kernels the
inequalities (\ref{ineqs}) are valid for both singlet and nonsinglet
combinations, as well as for gluon-gluon kernels for large $x$,
where the positivity of unpolarized NLO kernel is satisfied.

However, the quark-gluon and gluon-quark kernels
manifest tiny violations of (\ref{ineqs}) in the region of
large $x \sim 0.7$. This violation is by no means of
real importance for evolution, as it is completely
screened by the LO contribution satisfying positivity.

At the same time, this violation is of some theoretical interest.
Note that both these kernels contain only one $\gamma_5$ matrix
and are sensitive to its definitions while performing
$\varepsilon-$ regularization. Also, the interplay of polarized
and unpolarized kernels is very peculiar. Both contain
logarithmic and polynomial terms in $x$, not matching
each other and only (numerical) addition results in the small
violation of inequality (\ref{ineqs}).

One should note,
that the inequality analogous to (\ref{ineqs}) for the moments
of the splitting kernels (anomalous dimensions)
is, generally speaking,
not sufficient for positivity, since a
distribution with all  moments positive may be negative for small $x$.
Also, moments combine regular and singular at $x=1$ terms,
while only the first are essential for positivity.
The case discussed about, however, is free of singular terms and deals
with large $x$. Consequently, violation of positivity can be
seen in the moments \cite{Vog} (due to the smallness of the violation,
only the case of one of the two kernels is seen in the figures).

To conclude, we have found that NLO evolution
(when LO and LO kernels are added) preserves positivity.
The small violation for quark-gluon and gluon-quark kernels
may be related to the definition of $\gamma_5$ matrix.
This is under investigation

\newpage

\end{document}